\documentclass{optica-article}

\journal{opticajournal} % for journals or Optica Open

\articletype{Research Article}

\usepackage{lineno}
\usepackage{pdfpages}

\begin{document}

\title{Coronagraph-based wavefront sensors for the high Strehl regime}

\author{V. Chambouleyron,\authormark{1,*} J. K. Wallace,\authormark{2} R. Jensen-Clem,\authormark{1} B. Macintosh\authormark{1}}

\address{\authormark{1}Department of Astronomy \& Astrophysics, University of California, Santa Cruz, 1156 High Street, Santa Cruz, CA 95064, USA\\
\authormark{2}Jet Propulsion Laboratory, California Institute of Technology, 4800 Oak Grove Dr, Pasadena, CA 91109\\}

\email{\authormark{*}vchambou@ucsc.edu} %% email address is required; see note below about the corresponding author designation

% use {asbstract*} to suppress the copyright line. Copyright information will be added in production

\begin{abstract*}

A crucial component of the high-contrast instrumental chain in astronomy is the wavefront sensor (WFS). A key property of this component is its sensitivities, which reflect its ability to efficiently use incoming photons to encode the phase aberrations. 

This paper introduces a new class of highly sensitive wavefront sensors that approach the fundamental sensitivity limits dictated by physics. Assuming a high Strehl regime, we define what linear operator is describing the ideal WFS that would achieve maximum sensitivity. We then show that there is a substantial similarity between this ideal WFS and the second-order ideal coronagraph.

Leveraging the exhibited link between ideal wavefront sensing and coronagraphy, we propose a novel WFS concept based on high-performance coronagraphic architecture : the bivortex WFS. This sensor employs charge-2 vortex masks. Simulations for an ideal system demonstrate that this sensor achieves unprecedented sensitivity, even surpassing the highly sensitive Zernike WFS class (especially for low spatial frequencies), while paving the way for new high-contrast architectures integrating simultaneous sensing and coronagraphy.

\end{abstract*}

%%%%%%%%%%%%%%%%%%%%%%%%%%  body  %%%%%%%%%%%%%%%%%%%%%%%%%%
\section{Introduction}

High-contrast imaging systems are essential for observing faint astronomical objects close from bright sources by employing sophisticated techniques to suppress starlight down to extremely low levels. These systems integrate wavefront sensing and control with coronagraphy to control and eliminate starlight to achieve the necessary contrast for detailed observations such as direct imaging and characterization of exoplanets and/or circumstellar environments \cite{GPI,sphere}. Precise wavefront sensing and control is a crucial step as it prevents starlight leakage, which can compromise the quality of the science data by introducing unwanted noise in the final images \cite{faustine_ESO}.

A fundamental component of the wavefront sensing and control chain is the wavefront sensor (WFS), responsible for detecting aberrations and guiding corrections. One key property of a WFS is its sensitivity, which reflects its capability to efficiently utilize photons for encoding the phase of the incoming electromagnetic (EM) field. Enhanced sensitivity translates to the ability to obtain high-quality measurements with shorter integration times or fainter targets, allowing for instance increased sky coverage or faster wavefront corrections\cite{guyon2010,Chambouleyron_2023}.

In this paper, we propose a methodology to design highly sensitive WFSs suitable for high Strehl regimes (defined in this paper as Strehl-Ratio (SR) roughly above 80\%), applicable to both space-based missions or as second-stage WFSs in the case of ground-based observatories. The paper is organized as follow: in the section 2, we define the ideal WFS described by a straightforward operation on the input EM field which allows it to reach fundamental sensitivities within the small phase regime. We also highlight an important specificity of photon noise in WFS measurements. Section 3 explores the significant similarities between this ideal WFS and the second-order ideal coronagraph, which facilitates the development of practical implementation strategies and naturally leads to the introduction of a new WFS concept: the bivortex WFS.

\section{Building the ideal WFS in sensitivity}

\subsection{Framework for computing sensitivities in small phase regime}

We first outline the key findings from the mathematical framework introduced in \cite{Chambouleyron_2023}, which is employed throughout this paper. In the linear regime of a considered WFS, the variance of the phase estimation for a given mode $\phi_{i}$ with a number $N_{ph}$ of photons available for the measurements is expressed as follows:

\begin{equation}
\sigma^{2}_{\phi_{i}} = \frac{N_{sap}\times\sigma^{2}}{ s^{2}(\phi_{i})\times N_{ph}^{2}}+\frac{1}{s^{2}_{\gamma}(\phi_{i})\times N_{ph}}
\label{eq:propag_noise}
\end{equation}

the first term represents propagation of the uniform noise on the detector (typically dominated by read-out noise and dark current from the WFS detector) in the reconstruction, while the second term accounts for photon noise propagation. Here, $N_{sap}$ is the number of measurement points in the pupil and $\sigma^{2}$ denotes the variance of the uniform noise on the detector. The quantities $s$ and $s_{\gamma}$ represent the sensitivities to uniform and photon noise, respectively, for each mode $\phi_{i}$. These sensitivities can only take values between 0 (the WFS is not sensitive to the considered mode and only propagates noise) to 2 (indicating fundamental maximal sensitivity and minimal error propagation given the number of photons available \cite{paterson,PhysRevApplied.15.024047}). Thus we have:

\begin{equation}
\begin{split}
    0\leq\ &s \leq 2\\
    0\leq\ &s_{\gamma} \leq 2
\end{split}
\end{equation}

These quantities are crucial for comparing the performance of different WFSs in their linear regime and assessing how close they are to the fundamental WFS sensitivity limit. The sensitivity to uniform noise for each mode $\phi_{i}$ is computed using:

\begin{equation}
\label{sensibilité_mode_i}
s(\phi_{i}) = \sqrt{N_{sap}}\times\big|\big|\delta I(\phi_{i})\big|\big|_{2}
\end{equation}

\noindent where the term $\delta I(\phi_{i})$ is the entry of the interaction matrix (also called jacobian matrix, generally computed through a "push-pull" calibration) corresponding to the input mode $\phi_{i}$ (normalized in radian rms) and $||\cdot||_{2}$ is the 2-norm operator. The factor $\sqrt{N_{sap}}$ normalizes for the number of measurements points in a discrete framework, allowing for absolute sensitivity comparison between WFSs even with different samplings and can be discarded in the continuous case. To compute the photon noise sensitivity, because we are in small phase regime, the approximation is made that the distribution of the WFS measured intensities  $I(\phi)$ for a given phase $\phi$, which defines the structure of the photon noise, is dominated by the reference intensities $I_{0}$ corresponding to the measurements for the reference wavefront (assumed to be a flat wavefront in this study). In other words, we neglect the change $\Delta I(\phi)$ in the WFS intensities in presence of the phase $\phi$:

\begin{equation}
I(\phi) = I_{0} + \Delta I(\phi) \approx I_{0}
\end{equation}

\noindent Under this assumption, the main photon noise contribution is the map $\sqrt{I_{0}}$ and the photon noise sensitivity $s_{\gamma}$ can be computed as:

\begin{equation}
\label{sensibilité_mode_i_photon}
s_{\gamma}(\phi_{i}) = \bigg|\bigg|\frac{\delta I(\phi_{i})}{\sqrt{I_{0}}}\bigg|\bigg|_{2}
\end{equation}

\noindent where the division and the square root operations are defined element-wise.

\subsection{Harvesting the piston mode}

We present here a straightforward theoretical method for designing a WFS that achieves maximum sensitivity to both photon and uniform noise in the case of small phase regime. The electromagnetic (EM) field in the entrance pupil is represented as $\psi(\boldsymbol{r})$, where $\boldsymbol{r}$  denotes the position vector in the pupil plane. In this study, we assume the absence of amplitude errors and that the system operates around a flat wavefront. Accordingly, we can express:

\begin{equation}
\psi(\boldsymbol{r})  = P(\boldsymbol{r}) e^{i\phi(\boldsymbol{r}) }
\end{equation}

where $P$ is the binary aperture of the telescope (values at 1 for light collected by the telescope, 0 otherwise) and $\phi$ the phase (zero-mean and real) to be measured. Monochromatic light is assumed throughout the entire paper to simplify the formulation. For readability, the notation $\boldsymbol{r}$ is omitted for the rest of this section.\\

\textbf{Small Phase Regime}: As an initial and comprehensive approach, we assume the small phase regime. Under this approximation, the complex amplitude of the EM field can be expressed as $\psi = Pe^{i\phi} \approx P + iP\phi$. We introduce a WFS (hereafter referred to as the ideal WFS, denoted iWFS)  characterized by its linear operator $\mathcal{W}$ whose action on the EM field is defined as:

\begin{equation}
\mathcal{W}[\psi] = \mathcal{W}[P + iP\phi] = iP + iP\phi
\end{equation}

This implies that the iWFS introduces a phase shift of $\pi /2$ to the EM field "piston" mode relative to the rest of the EM field, which encodes the phase information. The intensities recorded by the iWFS can then be expressed as follows (noting that, since $P$ is the binary aperture, $|P|= P = P^{2}$):

\begin{equation}
I(\phi) = |\mathcal{W}[\psi]|^{2} = P + 2P\phi + P\phi^{2} \approx P + 2P\phi 
\label{eq:iWFS_small_phase}
\end{equation}

\noindent where the $\phi^{2}$ term can be neglected due to small-phase regime. Examining the structure of $I(\phi)$, it becomes apparent that the iWFS converts the variation in phase (in radians rms) into variation in intensities, with a scaling factor of 2. Notably, the derivative of the iWFS intensities with respect to the input phase can be easily computed (for positions where $ P(\boldsymbol{r})=1$):

\begin{equation}
\frac{\delta I}{\delta \phi} = 2
\end{equation}

Noting that  that the reference intensities $I_{0} = P$ for the iWFS and because $P$ was defined in a such way that $\sqrt{P} = P$, we can use equations \ref{sensibilité_mode_i} (in continuous framework) and \ref{sensibilité_mode_i_photon}, iWFS sensitivity for each position in the pupil where light is collected by the telescope is:

\begin{equation}
\begin{split}
    &s = 2\\
    &s_{\gamma} = 2
\end{split}
\end{equation}

As defined, the iWFS thus achieves the maximum possible sensitivities and represents the ideal WFS in terms of sensitivity in the small phase regime.\\

\textbf{General approach}: For a more realistic approach and to gain an understanding of the dynamic behavior of the iWFS, we define the iWFS operator $\mathcal{W}$ without assuming any constraints on the phase amplitude. The input EM field can be decomposed into the sum of  two quantities: a component $\mathcal{P}_{\parallel}$ co-linear with the piston mode and a component $\mathcal{P}_{\perp}$ orthogonal to it.

\begin{equation}
\psi = \mathcal{P}_{\parallel} + \mathcal{P}_{\perp}
\label{eq:decomposed}
\end{equation}

\noindent It is possible to write $\mathcal{P}_{\parallel} = \rho P$, where $\rho$ is the scalar projection of the EM field onto the piston mode: $\rho  = <P e^{i\phi}|P>$. The (complex) scalar $\rho$ is related to the SR $\mathcal{S}$ by the relation $|\rho| = \sqrt{\mathcal{S}}$ (see supplemental document  - section S1). The orthogonal component is expressed as $\mathcal{P}_{\perp} = \psi - \mathcal{P}_{\parallel} = Pe^{i\phi}- \rho P$. In the general case, the iWFS operator on the EM field shifts the EM projection on the piston mode $\mathcal{P}_{\parallel}$ by $\pi/2$, while leaving the orthogonal component $\mathcal{P}_{\perp}$ unaffected:

\begin{equation}
\begin{split}
\mathcal{W}[\psi] &= \mathcal{W}[\mathcal{P}_{\parallel} + \mathcal{P}_{\perp}] \\
&= i\mathcal{P}_{\parallel} + \mathcal{P}_{\perp} \\
 &= i\rho P + (Pe^{i\phi} - \rho P) \\
 &= Pe^{i\phi} + \rho P(i-1) \\
 &= Pe^{i\phi} + i\rho P\sqrt{2}e^{i\pi/4}
\end{split}
\label{eq:iWFS}
\end{equation}

\noindent By defining $\rho =|\rho |e^{i\chi}$, the intensities recorded by the iWFS are given by:

\begin{equation}
\begin{split}
I(\phi) &= |\mathcal{W}[\psi]|^{2} \\
&= P + 2|\rho |^{2}P - 2\text{Re}[iP^{2}e^{i\phi}|\rho |e^{-i\chi}\sqrt{2}e^{-i\pi/4}]\\
&= P + 2|\rho |^{2}P + |\rho |2\sqrt{2}P\sin\bigg(\phi-\frac{\pi}{4}-\chi\bigg)
\end{split}
\label{eq:iWFS_dynamics}
\end{equation}

\noindent This equation provides insight into the dynamics of the iWFS: (i) The phase signal is encapsulated within the sine function, which is indicative of the interferometric nature of the measurements. (ii) The offset term $\chi$ is actually negligible in the Maréchal approximation regime, roughly above $\mathcal{S}>0.3$, for which $\rho  = |\rho|$ (see  supplemental document section S1) so can be neglected assuming moderate/high Strehl regime. The $\pi/4$ offset reveals that the iWFS's linearity curve is asymmetric even in a high Strehl regime, a behavior akin to that observed in the Zernike WFS (ZWFS) \cite{ZeldaMamadou}. (iii) The factor $|\rho|$  preceding the sine term can be interpreted as a fringe contrast term. As the SR decreases, the sensitivity of the iWFS also diminishes, paralleling the behavior of the Zernike WFS or the non-modulated PWFS. It is worth noting that the SR defined here is "tip-tilt" sensitive. Consequently, the iWFS exhibits a rapid sensitivity drop for all spatial frequencies in the presence of a tip-tilt aberration (once again, similar to the ZWFS or the non-modulated PWFS). The iWFS is therefore only reaching high sensitivity in the case of high SR regime.

Additionally, by expanding the sine term  using the formula: $\sin(\phi-\frac{\pi}{4}-\chi) = \sin(\phi)\cos(\frac{\pi}{4}+\chi)-\cos(\phi)\sin(\frac{\pi}{4}+\chi)$ and assuming $\phi<<1$ (leading to $\rho \approx 1$ and $\chi \approx 0$), we logically recover the equation \ref{eq:iWFS_small_phase} for the small phase approximation derived earlier. 

\begin{equation}
I(\phi) = P + 2P + 2P(\phi - 1) = P + 2P\phi
\end{equation}

Overall, the iWFS is quite similar to the ZWFS, which is already known for operating close to fundamental sensitivity limits. However, while the ZWFS class exhibits reduced sensitivity for small spatial frequencies \cite{chambou_Z2WFS}, the iWFS maintains maximal sensitivity across all spatial frequencies. A this point, the iWFS remains a theoretical construct, but the section 3 will outline a pathway for its practical implementation.

\subsection{Energy in the reference wave}

The objective of this subsection is to highlight a fundamental difference in behavior between uniform noise and photon noise sensitivity. In equation \ref{eq:iWFS}, it can be observed that the iWFS signal is generated by the interference between two waves: the phase-shifted piston mode wave (referred to as the reference) and the wave carrying phase information. In the small phase approximation, the energy carried by the reference wave is significantly larger than that of the other wave. Reducing the amplitude of the reference wave will decrease the square root of the reference intensities proportionally to the signal, hence photon noise sensitivity remains unchanged (although sensitivity to uniform noise is reduced - see  supplemental document section S2 for further explanation). In other words, in a high Strehl regime, most of the photons are not effective for wavefront sensing in terms of photon noise sensitivity, as they contribute equally to both the signal and the photon noise. This shows that variations of the iWFS for which part of the reference wave energy is removed can still reach near optimally in terms of photon noise sensitivity, up to a limit defined in  supplemental document section S2. This difference in behavior between uniform noise and photon noise sensitivity can also be used to explain some of the properties of widely used class of highly sensitive WFS, specifically the pyramid and Zernike WFS. \\

\begin{figure}[h!]
\centering
        \includegraphics[width=0.7\columnwidth]{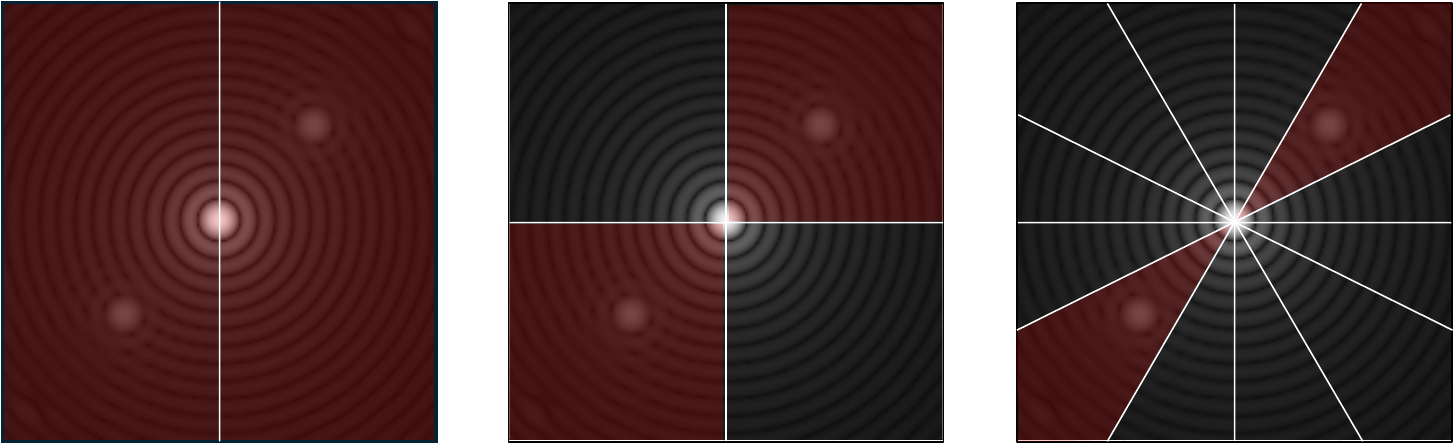}
    \caption{Part of the PSF useful for the sensing for the pyramid WFS class. The zone highlighted in red represents the PSF region interacting with the illustrated speckle. \textbf{Left :} rooftop prism, used for example in the case of the sharp Bi-O-edge. \cite{verinaud_biOedge} \textbf{Middle :} 4-sided PWFS. \textbf{Left :} 10-sides PWFS. In the small phase approximation where the PSF core carries most of the energy, all these sensors exhibit the same sensitivity to photon noise for the mode associated with the illustrated speckle.}
    \label{fig:PWFS}
\end{figure}

\textbf{Pyramid WFS class}: This effect elucidates a specific feature of the non-modulated pyramid WFS class. This class of WFS employs a pyramid-shaped mask to split the focal-plane EM field, resulting in the speckles interfering with the core of the point spread function (PSF) \cite{raga}. Regarding uniform noise, sensitivity decreases proportionally with the number of faces \cite{Chambouleyron_2023}. This can be easily understood as more faces induces more pupil images and therefore a greater spreading of the signal across the detector. However, for photon noise, all non-modulated pyramid WFS exhibit the same sensitivity : $s_{\gamma} = \sqrt{2}$ \cite{guyon2010,Chambouleyron_2023}. When examining the impact on a given spatial frequency (see figure \ref{fig:PWFS}), it is observed that adding more faces causes the speckle to interfere with a smaller portion of the PSF core. However, in a high Strehl regime, the PSF core contains significantly more energy than the speckle, making the increase in the number of faces to have only a marginal impact on photon noise sensitivity.\\

\textbf{Zernike WFS class}: The ZWFS class relies on a focal plane mask composed of a dimple, which serves to phase-shift the PSF core and thereby create the reference wave \cite{ZeldaMamadou}. Previous studies have shown that increasing the size of the dimple enhances sensitivity (for both uniform and photon noise) for high-spatial frequencies by adding more energy in the shifted part of the PSF \cite{chambou_Z2WFS}, albeit at the expense of sensitivity to low-spatial frequencies. One might question why the argument for the PWFS class does not apply to the ZWFS. Otherwise, photon noise sensitivity should remain unaffected by the amount of PSF energy located within the dimple. The main difference between the iWFS and the ZWFS class lies in the fact that, for the ZWFS, reducing the light passing through the dimple impacts both the reference wave and the second term of equation \ref{eq:iWFS}. For the iWFS, this would correspond to modifying the operator by introducing a real scalar $\alpha\in[0,1]$ such that:

\begin{equation}
\mathcal{W}[\psi] = \alpha \times i\rho P + (Pe^{i\phi} - \alpha \times \rho P)
\end{equation}

\noindent In that case, we can observe that decreasing $\alpha$ results in a significant increase of the energy carried by the EM wave $Pe^{i\phi} - \alpha \times \rho P$, making it much more energetic than the reference wave $\alpha \times i\rho P$ and therefore leading to a decrease in sensitivity.\\

In this section, the iWFS was introduced as a theoretical sensor that shifts the piston mode of the input EM field by $\pi /2$ to encode the phase in intensities. In the small phase approximation, the iWFS achieves maximal sensitivity across all spatial frequencies. Furthermore, it was also demonstrated that in the case of photon noise alone, most photons do not contribute effectively to optimal wavefront sensing as they contribute equally to both the signal and the noise. In the next section, we present a practical method to create an optical setup that approximates the iWFS by leveraging advancements in coronagraphy.

\section{Coronagraph-based WFS: a new class of extremely sensitive WFS}

\subsection{Parallel with coronagraphs: WFS eigenvalues decomposition}

\begin{figure}[h!]
\centering
        \includegraphics[width=0.85\columnwidth]{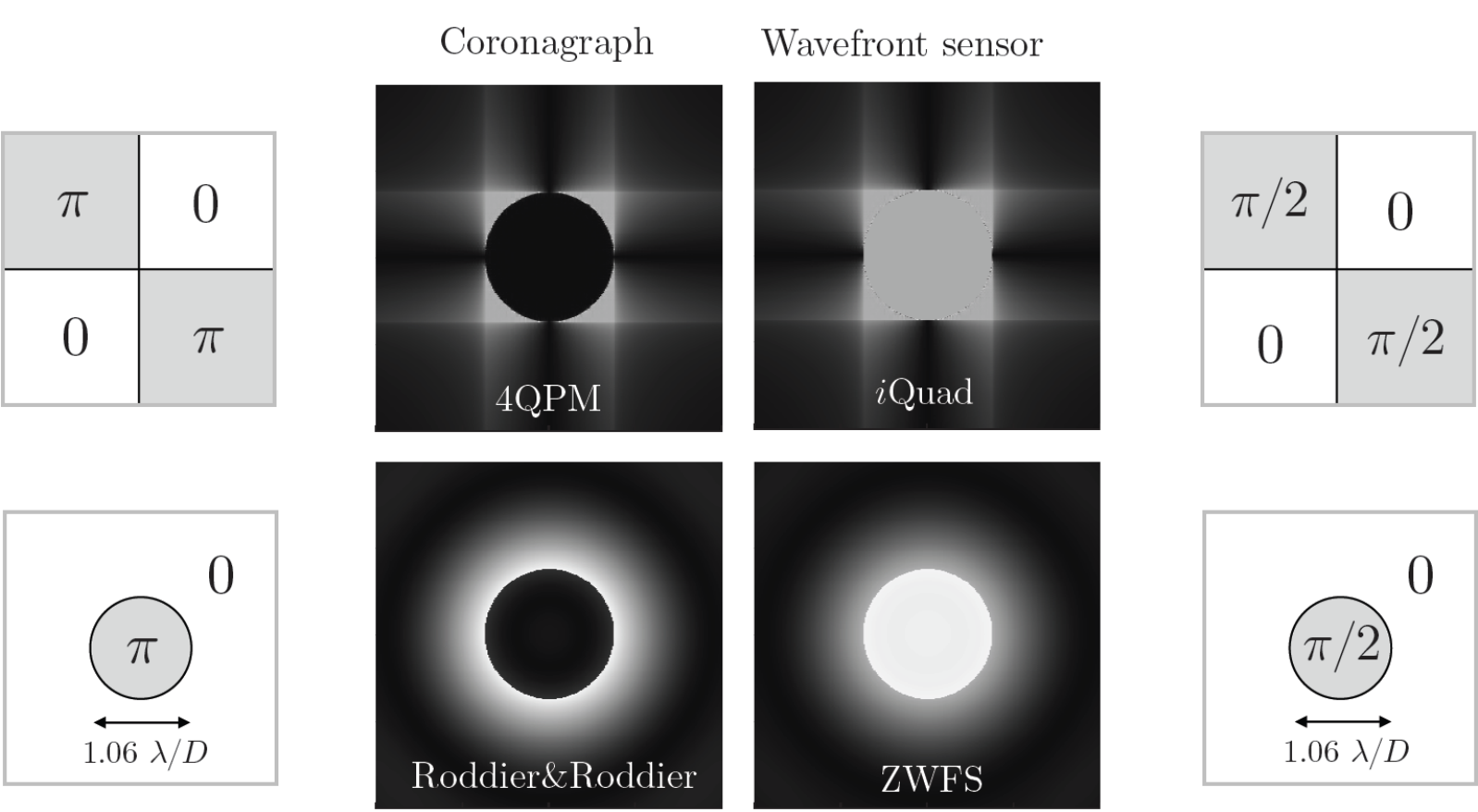}
    \caption{The $\pi$ versus $\pi/2$ phase shift: a common analogy between wavefront sensing and coronagraphy. \textbf{Top-left :} Phase of the 4 quadrant mask (4QPM) coronagraph and associated intensities in the pupil plane. \textbf{Top-right :} Same but for the WFS version of it, the so-called $i$Quad. \textbf{Bottom-left and botton-right:} Same analogy with the Roddier \& Roddier coronagraph, consisting of a circular $1.06\ \lambda/D$ diameter dimple inducing a $\pi$ phase-shift on the PSF core and the associated WFS: the ZWFS.} 
    \label{fig:wfs_corono}
\end{figure}

The connection between wavefront sensing and coronagraphy is often presented through the fact that coronagraphs create a reference wave shifted by $\pi$ to induce destructive interferences in a pupil plane, while wavefront sensors apply a $\pi/2$ shift to produce constructive interferences. Two notable examples of this concept are the Four Quadrant Phase Mask (4QPM) \cite{4QPM} and the Roddier \& Roddier \cite{Roddier_1997} coronagraphs, which can be adapted into wavefront sensors by modifying the phase shift they induce on the electromagnetic (EM) field. Figure \ref{fig:wfs_corono} illustrates the phase of the focal plane masks and the Lyot stop plane images (for a flat wavefront) of these two coronagraphs (left) and their corresponding wavefront sensor variants : the iQuad \cite{fauv_iquad} and the ZWFS \cite{ZeldaMamadou}, respectively (right).

\begin{figure}[h!]
\centering
        \includegraphics[width=1\columnwidth]{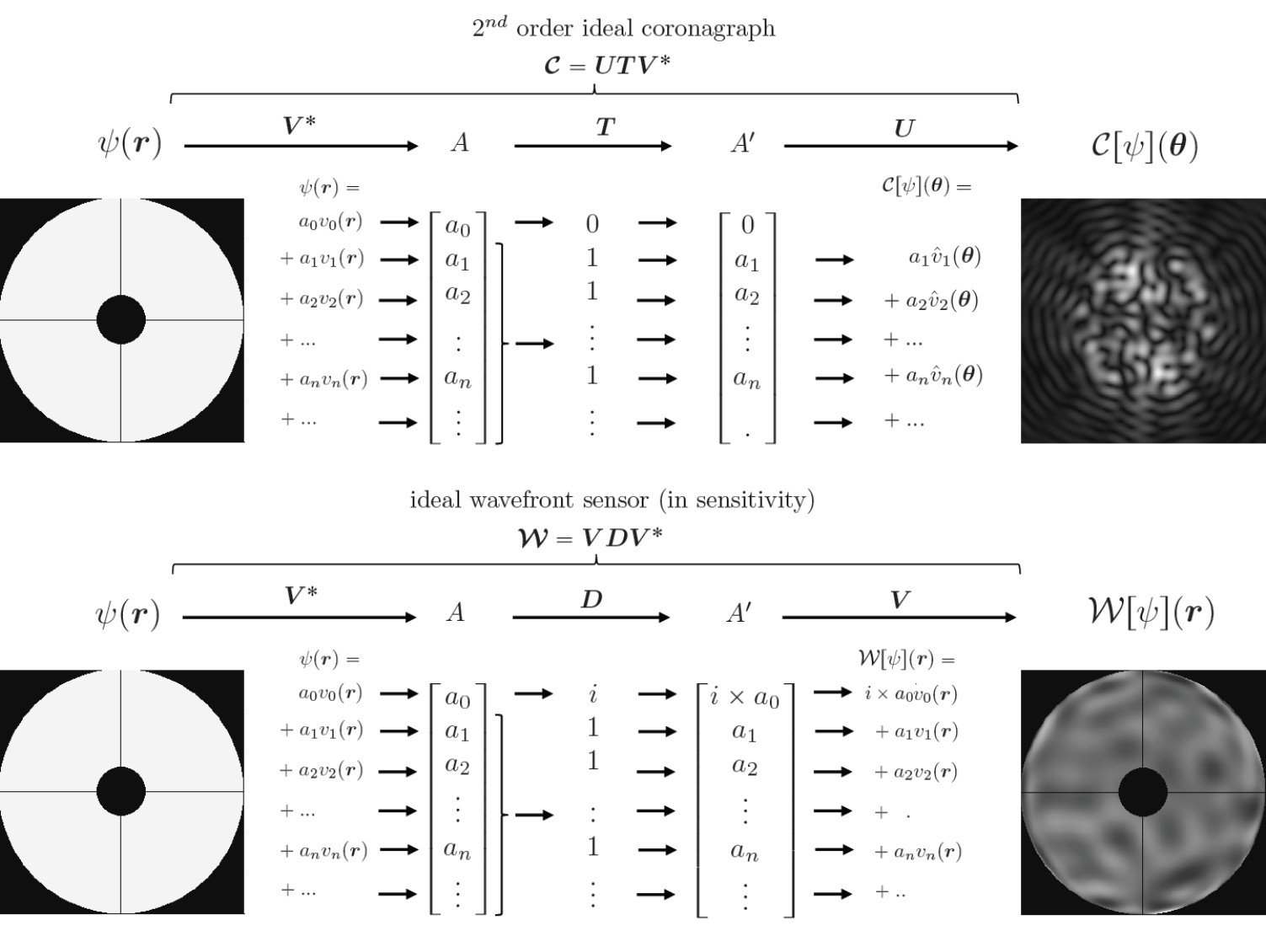}
    \caption{Similarities between the operators of the second-order ideal coronagraph and the ideal WFS. $\{v_{n}(\boldsymbol{r})\}$ is an orthornormal basis of $\psi(\boldsymbol{r})$ where $v_{0}(\boldsymbol{r})$ is the EM piston mode. EM fields are represented in absolute value. Position vector $\boldsymbol{r}$ is used for pupil plane, while $\boldsymbol{\theta}$ is used for focal plane. Figure inspired by \cite{belikov_2021}.}
    \label{fig:svd_optical}
\end{figure}

We propose to extend the coronagraph/WFS analogy further. It was demonstrated in the previous section that the iWFS operates on the EM field by shifting the piston mode by $\pi/2$. This operation is closely related to the function of the second-order ideal coronagraph, which aims to remove the EM field piston mode to fully reject on-axis starlight in the case of an unresolved star. As described in \cite{Guyon_2006} and \cite{belikov_2021}, the second-order ideal coronagraph operator on the EM field can be represented through its singular values decomposition. This representation is illustrated in the top part of figure \ref{fig:svd_optical} (inspired by \cite{belikov_2021}), where one can describes the second-order ideal coronagraph operator as a linear combination of passive (energy is conserved or reduced) optical operators, chosen as described in \cite{belikov_2021}:
\begin{enumerate}
    \item Choosing a set of orthonormal modes $\{v_{n}(\boldsymbol{r})\}$ that represents the noise (the starlight on axis, \textit{i.e} the piston mode in the case of the second-order ideal coronagraph) and the signal (all off-axis sources). The input EM field $\psi$ can then be represented as:
    
    \begin{equation}
\psi(\boldsymbol{r}) = \sum_{n=0}^{\infty} a_{n} v_{n}(\boldsymbol{r})
\end{equation}

using notation from previous section, we choose $v_{0}(\boldsymbol{r}) = P(\boldsymbol{r})$ the piston mode (and so $a_{0} = \rho$). This decomposition is represented by a matrix $\boldsymbol{V}^{*}$ (this operator can be interpreted as a "mode-sorting" operator) where $\cdot^{*}$ is the conjugate transpose.
    \item Filtering the unwanted modes, in the case of the second-order ideal coronagraph that means attenuating to 0 the piston mode  and leaving untouched the other modes. This can be represented by a diagonal matrix $\boldsymbol{T}$ performing the "modes selection".
    \item Finally, forming an image of the sky in science plane. Generally this operation can be seen as performing the reverse "modes-sorting" operation $\boldsymbol{V}^{*}$ and then forming an image : in that case it can be described as the operator $\boldsymbol{U} = \boldsymbol{\mathcal{F}}\boldsymbol{V}$, where $\boldsymbol{\mathcal{F}}$ is the Fourier transform matrix.
\end{enumerate}

We can apply exactly the same idea to the iWFS concept. To build the iWFS passive linear operator, it is possible to pursue the following steps (bottom figure \ref{fig:svd_optical}):

\begin{enumerate}
    \item Choosing a set of orthonormal modes ${v_{n}(\boldsymbol{r})}$ that represents the piston mode and the rest of the signal (represented by matrix $\boldsymbol{V}^{*}$).
    \item The piston mode is shifted by $\pi/2$ with respect to the other modes. This can be represented by a diagonal matrix $\boldsymbol{D}$ performing this "piston shift". The elements in this diagonal matrix are the WFS eigenvalues.
    \item Finally, we apply the inverse transformation of the first step by applying $\boldsymbol{V}$. The resulting EM field is therefore:
        \begin{equation}
\mathcal{W}[\psi](\boldsymbol{r}) = i\times a_{0}v_{0}(\boldsymbol{r}) + \sum_{n=1}^{\infty} a_{n} v_{n}(\boldsymbol{r})
\end{equation}
\end{enumerate}

To summarize, the second-ideal coronagraph passive linear operator can be written $\boldsymbol{\mathcal{C}} = \boldsymbol{U}\boldsymbol{T}\boldsymbol{V}^{*}$ and will generally take the form $\boldsymbol{\mathcal{C}} = \boldsymbol{\mathcal{F}}\boldsymbol{V}\boldsymbol{T}\boldsymbol{V}^{*}$, while the iWFS operator can be written $\boldsymbol{\mathcal{W}} = \boldsymbol{V}\boldsymbol{D}\boldsymbol{V}^{*}$ . This demonstrates the fundamental similarity between the second-order ideal coronagraph and the iWFS: the former cancels the piston mode, while the latter phase-shifts it by $\pi/2$. It is important to point out that the design of the iWFS is less stringent than that of a second-order ideal coronagraph. Indeed, a small leakage up to a few percent of the piston mode into other modes would not significantly affect the iWFS's sensitivity, although it would be highly detrimental for the coronagraph (for a quantitative analysis of the impact of piston mode leakage on sensitivities, see supplemental document section S3). Previous studies have shown that constructing a second-order ideal coronagraph, though requiring complex engineering beyond the current state of the art, is feasible \cite{Guyon_2006,belikov_2021}. By applying the same reasoning, it is possible to develop the iWFS, albeit at the cost of a similarly complex engineering design. These similarities naturally suggest leveraging the architecture of high-performing coronagraphs to create a WFS that approaches the iWFS.

\subsection{Implementation with Fourier-filtering optical layouts}

\begin{figure}[h!]
\centering
        \includegraphics[width=1\columnwidth]{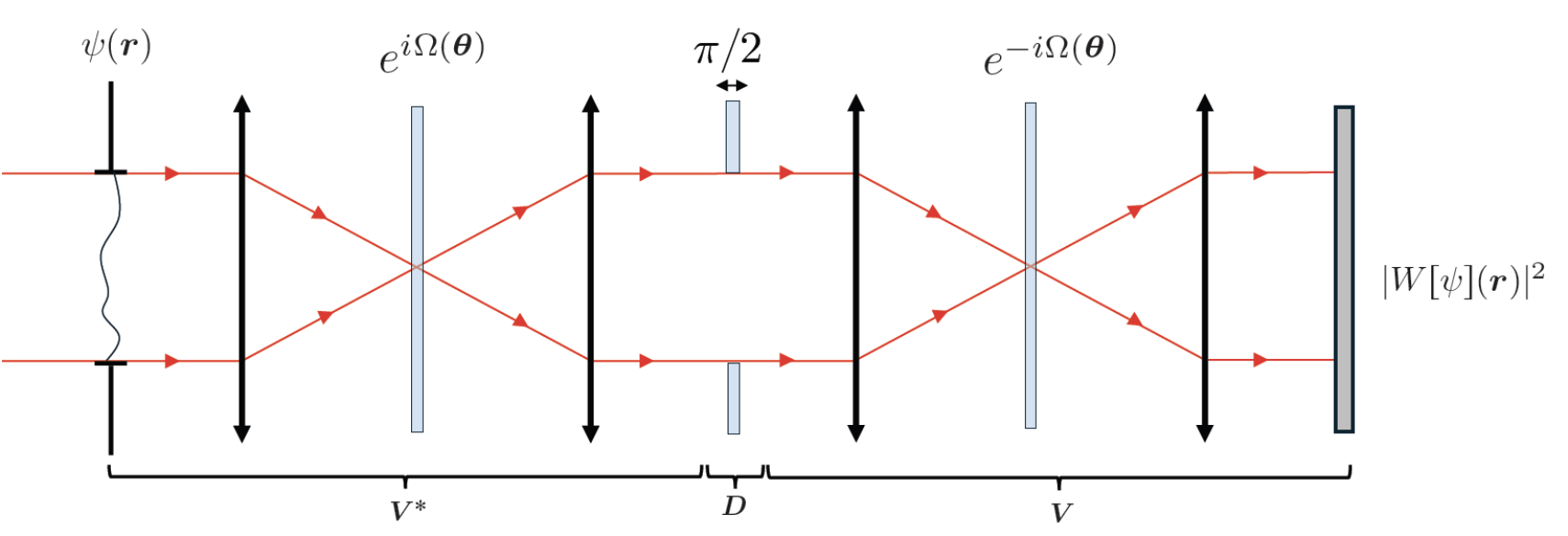}
    \caption{The biFFWFS concept: approaching the iWFS architecture described figure \ref{fig:svd_optical} with Fourier-filtering optical schemes. The first Fourier-filtering stage acts as the modes-sorting matrix $\boldsymbol{V^{*}}$, the phase-shifting Lyot as the matrix $\boldsymbol{D}$ and the second Fourier-filtering stage is reverting the operation performed by the first stage, hence playing the role of $\boldsymbol{V}$.} 
    \label{fig:layout}
\end{figure}

By leveraging the similarities between coronagraphs and WFS discussed in the previous subsection, a practical approach to implementing a WFS that approximates the iWFS is to use a Fourier-filtering architecture. In coronagraphy, such an architecture is widely used by introducing a focal plane mask designed to reject light from the piston mode outside the pupil footprint. For a simple coronagraph design (with no pupil plane or Fresnel plane apodization), this filtering operation can be assimilated to the operator $\boldsymbol{V}^{*}$ introduced previously. In a conventional coronagraph, the rejected starlight is managed with an opaque mask known as the Lyot stop, which functions as the operator $\boldsymbol{T}$ (figure \ref{fig:svd_optical}). Following the concept illustrated in figure \ref{fig:svd_optical}, we propose replacing the Lyot stop with a "phase-shifting Lyot" that introduces a $\pi/2$ phase shift to the rejected light. After this operation, the objective is to redirect all the light from the Lyot plane back into the pupil footprint. This can be achieved by performing the inverse operation of the initial Fourier-filtering stage. For a fully transmissive coronagraphic focal plane mask, this operation is straightforwardly accomplished using the reverse (conjugate) of the focal plane mask. An example of such a layout is illustrated in figure \ref{fig:layout}: the focal plane mask is fully transmissive and defined by the function $e^{i\Omega(\boldsymbol{\theta})}$, where $\Omega$ is a real function and $\boldsymbol{\theta}$ the position vector in the focal plane. The rejected light in the Lyot pupil plane is phase-shifted by $\pi/2$ and then passes through another Fourier-filtering stage with the conjugate mask: $e^{-i\Omega(\boldsymbol{\theta})}$. This new kind architecture defines a new class of WFS : the bi-Fourier-Filtering WFS class (biFFWFS).\\

\subsubsection*{The bivortex WFS} 

\begin{figure}[h!]
\centering
        \includegraphics[width=.95\columnwidth]{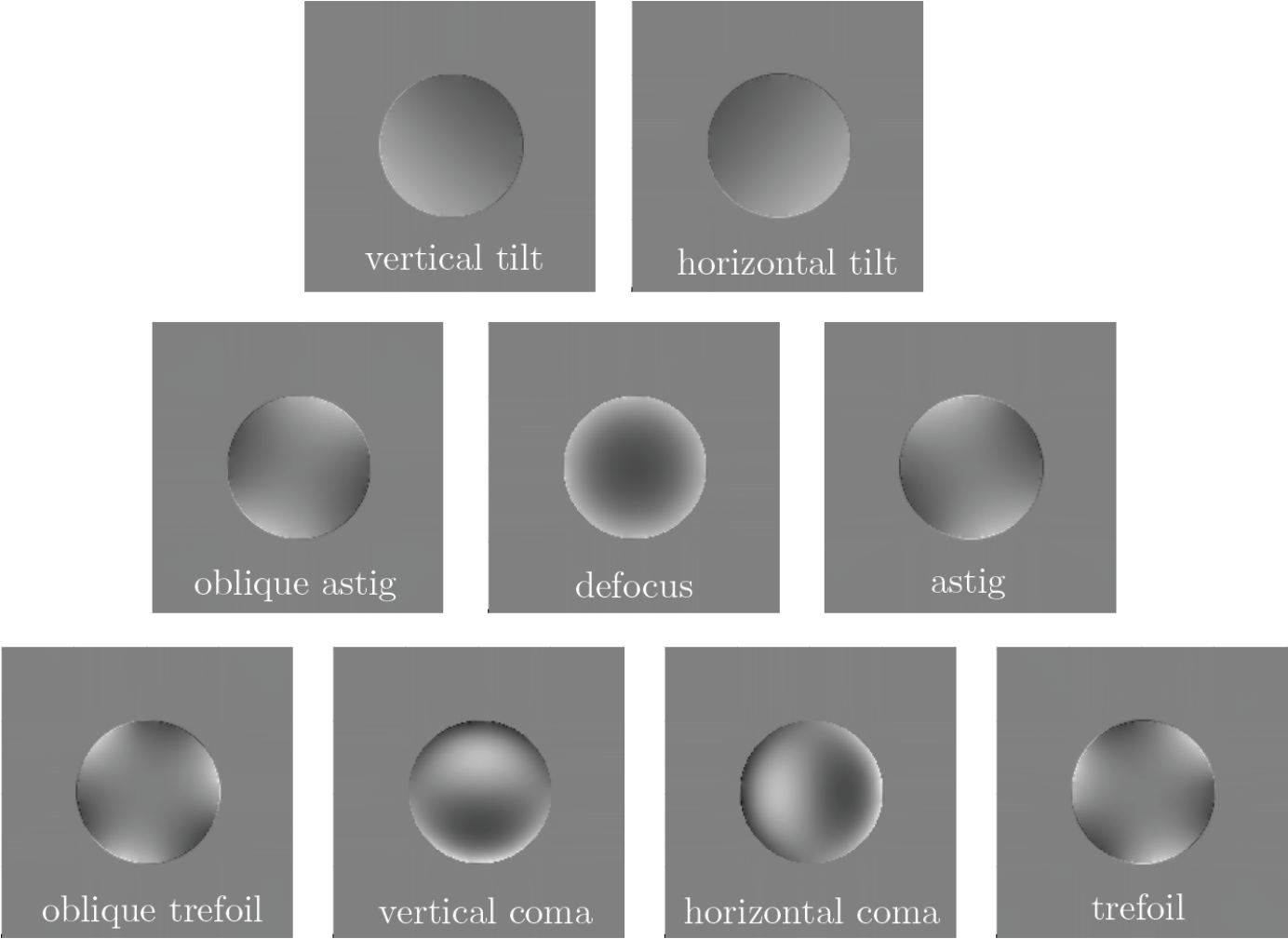}
    \caption{Linear response of the bivortex WFS to the first 9 Zernike polynomials (excluding piston mode). The plotted map represent the reference-removed response to small Zernike aberration $\Delta I(Z_{i}) = I(Z_{i})-I_{0}$, where the $I_{0}$ map corresponds to the measurement for a flat wavefront.}
    \label{fig:bivortex_response}
\end{figure}

For a full circular aperture, a well-known mask that approximates the second-order ideal coronagraph is the charge-2 vortex coronagraph \cite{vortex_foo}. By writing the focal plane position vector $\boldsymbol{\theta} = [\theta_{x},\theta_{y}]$, the phase function of this mask is defined by $\Omega(\theta_{x},\theta_{y}) = 2 \times \tan^{-1}(\theta_{y}/\theta_{x})$. As a direct application of the concepts introduced in this paper, we propose a new WFS concept: the bivortex WFS. This WFS design utilizes a charge-2 vortex mask as the coronagraphic focal plane mask and its conjugate (reverse mask) for the second focal plane mask. The Lyot stop plane is replaced by a phase-shifting Lyot, where the rejected light is shifted by $\pi/2$ while the light in the pupil footprint remains unaltered. In this case of an ideal full circular aperture, the bivortex WFS measurements for a flat wavefront give a map similar to the ones obtained when recording a simple pupil image. The simulated linear responses of this new WFS to the first Zernike modes is illustrated in figure \ref{fig:bivortex_response}. As anticipated, the bivortex WFS behaves like a phase sensor due to its measurement principle, which is designed to closely approximates the behavior of the iWFS described in equation \ref{eq:iWFS}. For low-order modes (such as tip-tilt and astigmatism), the bivortex WFS induces a rotation of the intensity pattern relative to the incoming phase. \\

\begin{figure}[h!]
\centering
        \includegraphics[width=0.9\columnwidth]{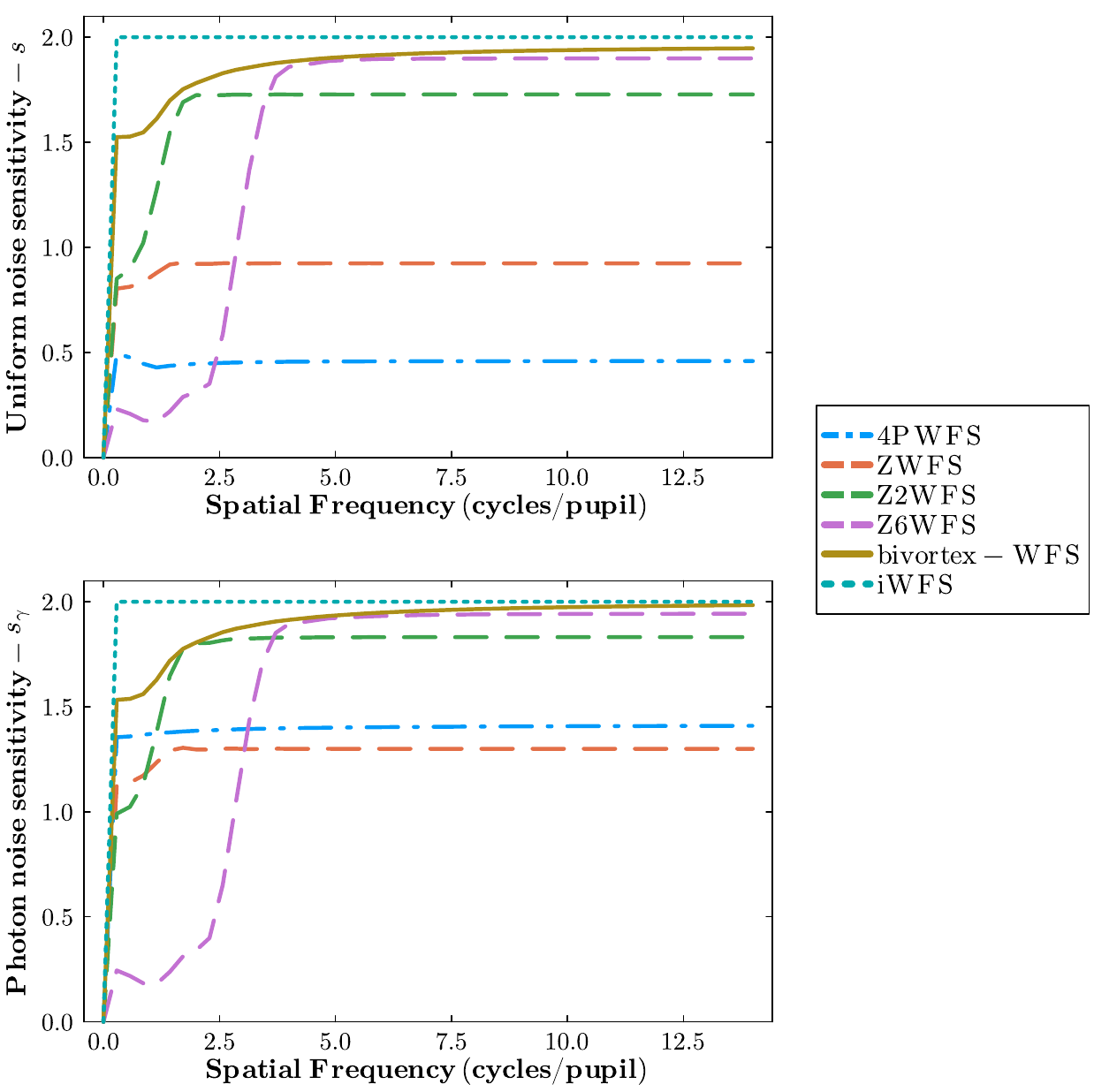}
    \caption{Sensitivity curves for different WFS, including the bivortex WFS introduced in this paper (4PWFS is non-modulated). \textbf{Top :} Sensitivities for uniform noise. \textbf{Bottom :} Sensitivities for photon noise.}
    \label{fig:sensitivities}
\end{figure}

\textbf{Sensitivities}: Sensitivities of the bivortex WFS for both uniform noise and photon noise can be computed according to equations \ref{sensibilité_mode_i} and  \ref{sensibilité_mode_i_photon}. We evaluate its sensitivity in comparison with five other WFS types: the non-modulated four sided PWFS (4PWFS), the ZWFS ($1\lambda/D$ dimple diameter), the Z2WFS (ZWFS with a $2\lambda/D$ dimple diameter), the Z6WFS (ZWFS with a $6\lambda/D$ dimple diameter) and the iWFS as defined in the previous section. For our comparison, we computed sensitivities with respect to a set of Fourier modes. The sensitivity for a given spatial frequency is calculated as the quadratic mean of the sensitivities for the sine and cosine components. Results are depicted in figure \ref{fig:sensitivities}, where all curves present a sharp discontinuity at 0 cycles/pupil which corresponds to the immeasurable global phase piston mode. For the ZWFS class, the trade-off between sensitivity for low and high spatial frequencies is clearly visible \cite{chambou_Z2WFS}. However, the bivortex WFS doesn't have this limitation and outperforms all the WFS presented here across all spatial frequencies. It is worth mentioning that the optimized versions of the ZWFS, which is not presented here, consists of adapting the shape of the mask to the PSF \cite{chambou_SPIE2022,landman2022} (or inversely adapting the pupil shape to the mask \cite{haffert_PIAA_zernike}). In that case, the resulting WFSs can be approximated by achieving the best sensitivities across all ZWFS curves, but remains slightly less sensitive than the bivortex WFS for low-order modes. For a full circular aperture, the bivortex WFS is, to our knowledge, the most sensitive WFS ever proposed. For a full circular aperture, the bivortex WFS is, to our knowledge, the most sensitive WFS ever proposed. It is also important to highlight another major advantage of the bivortex WFS over the ZWFS class which concerns chromaticity. Both the ZWFS and the vortex masks can impart achromatic phase shifts using liquid crystals that induce phase delay through geometric phase \cite{Mawet:09,Doelman:19}. However, even with an achromatic phase shift, the ZWFS’s response still varies with wavelength due to the scaling of diffraction with wavelength versus its fixed dimple diameter. In contrast, the vortex masks inside the bivortex WFS have no radial structure, allowing the sensor to remain achromatic over a large bandwidth.\\

\begin{figure}[h!]
\centering
        \includegraphics[width=0.7\columnwidth]{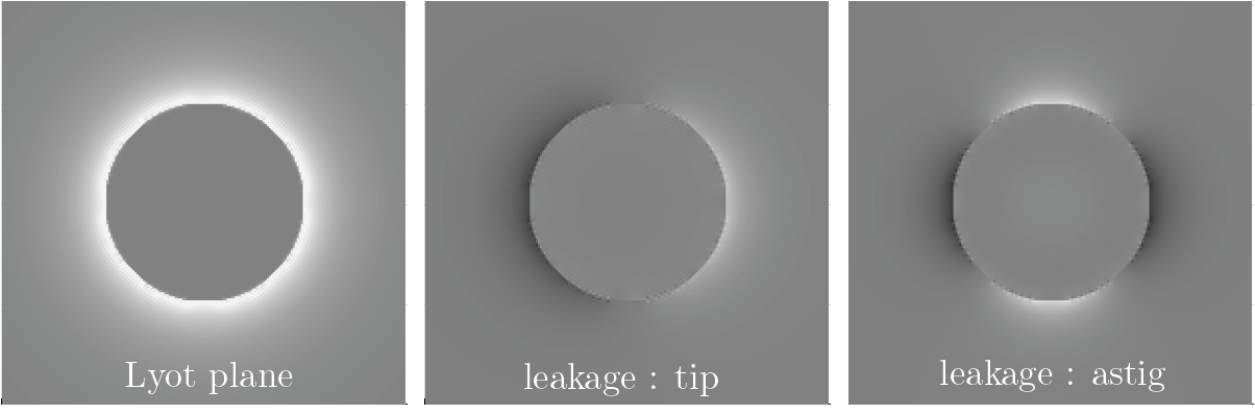}
    \caption{Leakage of low-order modes in the Lyot stop plane of a charge-2 vortex. \textbf{Left :} Intensities for a flat wavefront.\textbf{Middle :} Reference-removed intensities in presence of a tip. \textbf{Left :}  Reference-removed intensities in presence of an astigmatism. These leakage terms illustrate why the bivortex WFS is not reaching optimal performance for low-spatial frequencies.}
    \label{fig:leakage}
\end{figure}

The bivortex WFS does not achieve theoretical maximum sensitivities for low-order spatial frequencies because the charge-2 vortex coronagraph is not a perfect second-order ideal coronagraph, even for an idealized full circular pupil, as it exhibits a non-optimal throughput curve \cite{Guyon_2006}. As a result, part of the energy carried by the low-order spatial frequencies mixes with the reference wave and is not sensed anymore, causing a slight drop in its sensitivities. This effect is illustrated in figure \ref{fig:leakage}, which shows light leakage outside of the pupil footprint in the Lyot plane of small tip-tilt and astigmatism aberrations before a charge-2 vortex coronagraph.\\

\begin{figure}[h!]
\centering
        \includegraphics[width=1\columnwidth]{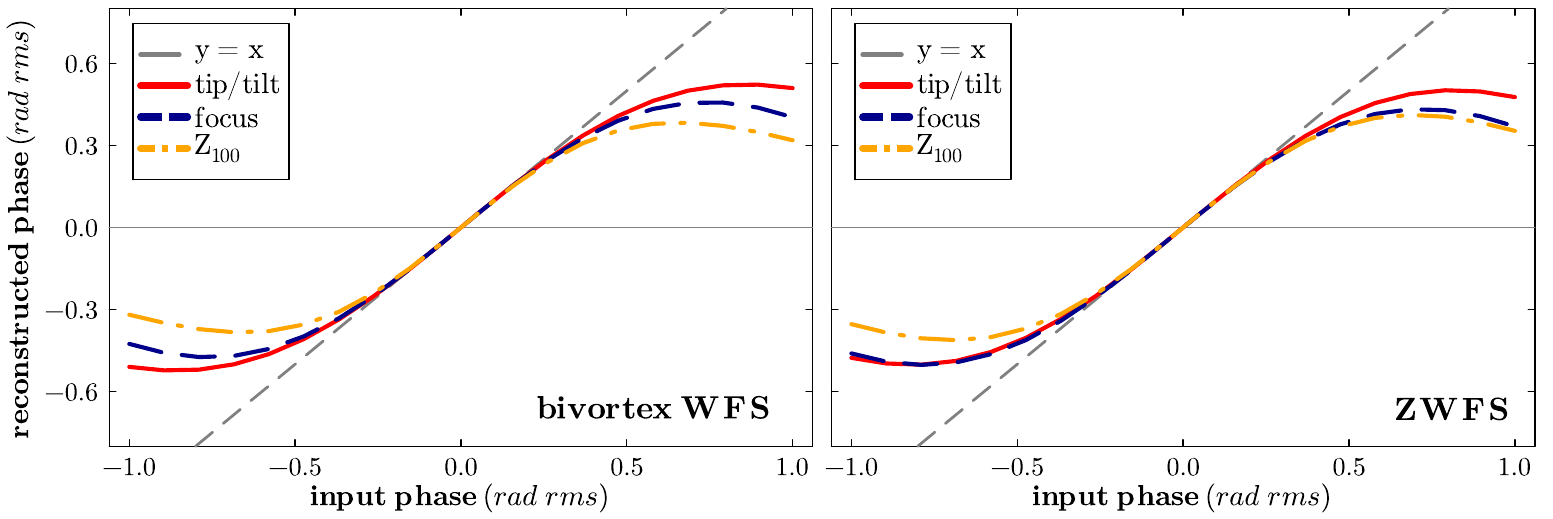}
    \caption{Linearity curves for tip//tilt, focus and high-order Zernike mode ($Z_{100}$). \textbf{Left:} bivortex WFS.\textbf{Right:} ZWFS. Dynamic range is extremely similar between these two WFSs.}
    \label{fig:linearity}
\end{figure}

\textbf{Dynamic range}: A preliminary evaluation of the bivortex WFS dynamic range was conducted in simulations by generating linearity curves for Zernike modes. These curves were obtained by injecting an input mode with increasing amplitude and reconstructing the phase linearly using the inverse of the interaction matrix. The results are presented in figure \ref{fig:linearity}, alongside the linearity curves for the ZWFS (with a dimple diameter of $1 \lambda /D$) for comparison. For these curves, only tip/tilt, focus and a high-order Zernike mode ($Z_{100}$) are plotted. The behavior of the two sensors is remarkably similar, mainly because their dynamic range is primarily constrained by the interferometric nature of their measurements, which closely resembles the characteristics of the iWFS, as described by equation \ref{eq:iWFS_dynamics}. This indicates that the bivortex WFS is expected to perform in a manner resembling that of the ZWFS. Importantly, it also demonstrate that the higher sensitivity of the bivortex WFS is achieved through a more effective utilization of photons, without incurring any penalties regarding dynamic range.\\

 To conclude this section in terms of potential applications, the bivortex WFS could for instance be employed as a second-stage WFS in extreme AO systems feeding high-contrast imaging instruments \cite{PCS_kasper}. Its increased sensitivity, especially for low-spatial frequencies, would enable faster and, therefore, more effective correction of atmospheric distortions near the star, enhancing the potential to detect exoplanets at small angular separations. On top of that, its capacity to sense directly the phase would be a strong advantage to measure telescope-induced aberrations such as segment phasing \cite{salama2024}, low-wind-effect \cite{Milli_2018} or even petal modes \cite{bertrou2022}. However, it is essential to recognize that the simulations were conducted assuming an ideal optical setup. To fully assess performance of this new WFS, it will be necessary to examine additional realistic factors that are out of the scope of this paper, such as the finite dimensions of the phase-shifting Lyot (whose inner radius may need to be slightly oversized to mitigate pupil edge effects), more realistic pupil shapes, mismatches between the mask and its inverse, chromatic effects, dynamic range for different aberrations in realistic operation regimes, and other practical considerations.

\subsection{Combining coronagraphic and wavefront sensing capabilities}

\begin{figure}[h!]
\centering
        \includegraphics[width=0.7\columnwidth]{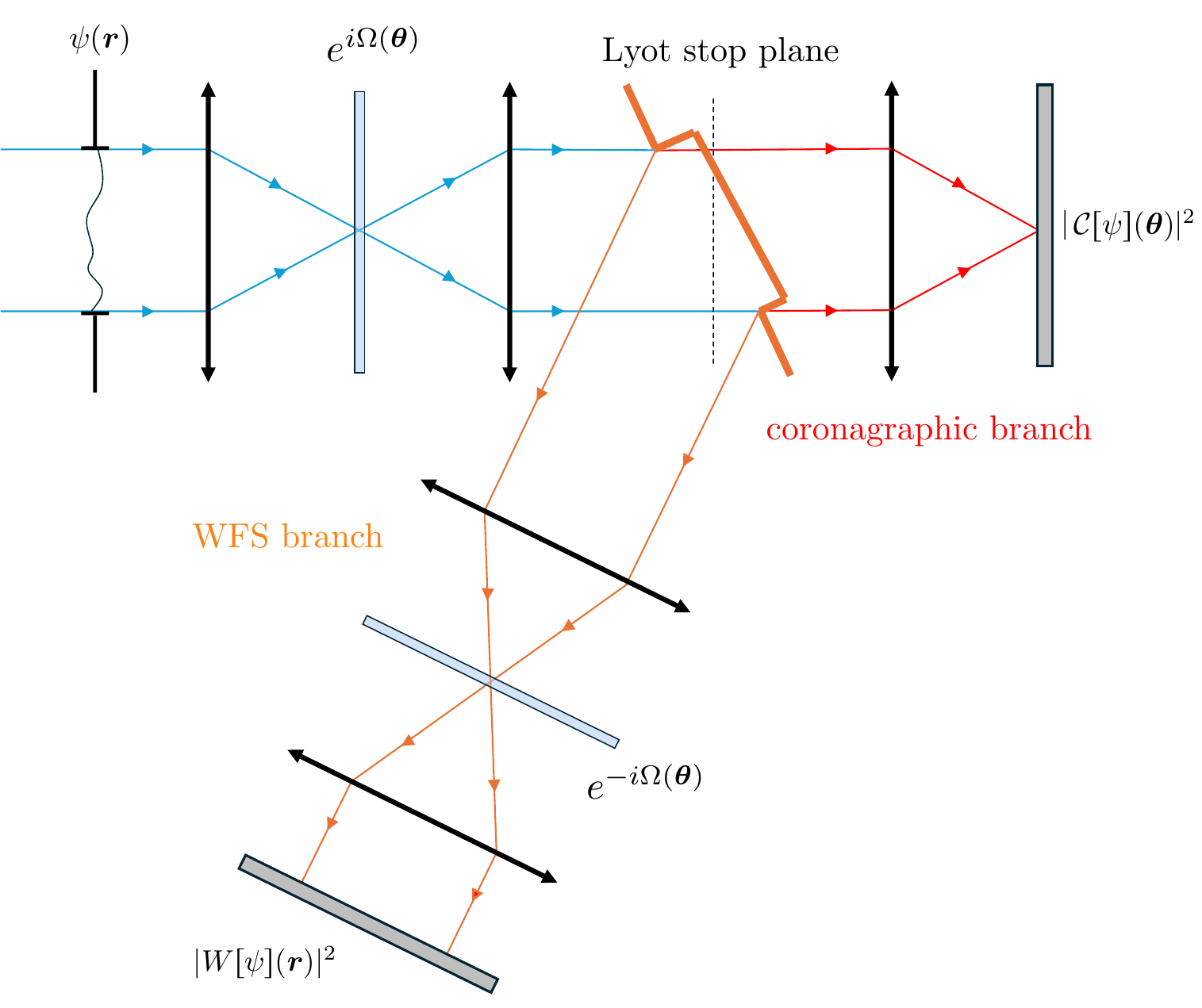}
    \caption{ A dual setup leveraging the similarities between wavefront sensing and coronagraphy to perform both at the same time. The same first mask is used for both the coronagraph and the WFS branch.}
    \label{fig:coro_WFS}
\end{figure}

In addition to providing extremely sensitive WFS architecture, such as with the bivortex WFS, the coronagraph-based WFS approach and its implementation through the biFFWFS configuration naturally pave the way for developing configurations that integrate both coronagraphic and wavefront sensing functions. This approach allows for the use of the same focal plane mask for both coronagraphy and WFS, drastically reducing the non-common path aberrations which represents one of the main contrast limitation in current high-contrast instruments \cite{faustine_ESO}. A proposition to implement such a dual setup is given figure \ref{fig:coro_WFS}. The light is focalised on a first mask which is used as the coronagraphic focal plane mask as well as the first part of the biFFWFS configuration. In the following pupil plane, light is split between two branches. One part of the light sees a Lyot stop and is then directed to a focal plane science camera: it is the coronagraphic branch. The other part reaches a phase-shifting Lyot (in reflection in the schematic figure  \ref{fig:coro_WFS}) and then passes through the reverse focal plane mask before reaching a pupil plane camera: it is the WFS branch. The splitting at the Lyot plane can be achieved using either a dichroic for out-of-band sensing or a beamsplitter for in-band sensing.

However, this architecture introduces a significant challenge: while the iWFS can be approximated using the second-order ideal coronagraph ability to extract the piston mode from the input EM field, this kind of coronagraphs is often not used in practice due to its lack of robustness with respect to low-order aberrations (especially tip/tilt) and stellar angular diameter \cite{Guyon_2006}. To address this lack of robustness, higher-order coronagraphs are employed in practical implementations. The objective of these coronagraphs is to reject a greater number of modes beyond just the piston mode. In the case of vortex coronagraphs, the generation of such high-order configurations can be achieved by increasing the charge of the vortex mask. In relation to the WFS design proposed in this paper, this involves incorporating additional modes into the reference wave, aside from the piston mode. As a result, these additional modes remain unmeasured, as they are not phase-shifted with respect to the piston mode. The design of the Lyot phase-shifting device must be adapted to address this issue and mitigate possible significant sensitivity losses for these additional rejected modes. One way could be to oversize the phase-shifting Lyot with respect to the pupil, to prevent the rejected modes from being phase-shifted along with the piston mode, but such a design will necessarily come at the expense of the sensitivities. The challenge of designing a phase-shifting Lyot in order to cope with high-order coronagraph masks will be explored in a more extensive study on the bivortex WFS, which will include an assessment of how vortex charge impacts sensing performance.

%Higher-order coronagraphs, which are used instead, can introduce other modes besides the piston mode into the reference wave.

\section{Conclusion}

After reviewing the two types of sensitivities that describe a WFS's efficiency in encoding phase information into intensities, and how these can be computed under the small phase regime assumption, we theoretically defined the optimal WFS operator on the EM field to maximize sensitivity in this regime. This ideal WFS, referred to as the iWFS, involves phase-shifting the input EM field's projection onto the EM piston mode by $\pi/2$, leaving the rest of the field unchanged. This process creates interferences that optimally encode the phase, assuming a high Strehl regime. The iWFS closely mimics the second-order ideal coronagraph behvior, which role is to suppress the EM field's projection onto the piston mode. Thus, we have shown that wavefront sensing and coronagraphy share therefore fundamental similarities and face a same challenge: extracting the piston mode from the rest of the EM field to perform specific operations on it.

Building on this concept, we proposed a novel WFS based on a high-performance coronagraph architecture: the bivortex WFS. This sensor comprises two cascaded Fourier-filtering stages, with the first stage featuring a charge-2 vortex and the second being its conjugate. The intermediate pupil plane incorporates what we refer to as a phase-shifting Lyot, which induces a $\pi/2$ phase shift on the piston mode. We demonstrated that the bivortex WFS approaches the sensitivity of the ideal WFS and achieves the highest simulated sensitivity for a WFS in ideal conditions with a full circular aperture. Notably, the bivortex WFS surpasses the ZWFS class, especially at low spatial frequencies, which are critical for high-contrast imaging at small angular separations. Such a sensor could be constructed to perform broadband by using liquid crystal vortices, which are already widely utilized within the high-contrast imaging community \cite{Mawet:09}. An implementation using such liquid crystals would also enable the development of a vector bivortex WFS, capable of measuring not only the phase but also the amplitude of the electromagnetic field, as proposed with the vector ZWFS \cite{Doelman:19}.

Only the ideal and basic concept of the bivortex WFS has been introduced in this paper. Further realistic simulations and studies are needed to fully evaluate the potential of this new WFS and address possible limitations. An important area for further analysis is integrating WFS capability with coronagraphic functionality, particularly using higher charge vortex masks. Regardless of the bivortex concept, the strong link highlighted in this paper between the ideal WFS and the field of coronagraphy suggests that developments in coronagraphy, especially those involving photonics \cite{2024_photonics,quantum_corono}, can be applied to designing WFSs that approach the fundamental sensitivity limits.

\section*{Acknowledgments}       
 
This work benefited from the 2024 Exoplanet Summer Program in the Other Worlds Laboratory (OWL) at the University of California, Santa Cruz, a program funded by the Heising-Simons Foundation and NASA. The authors would also like to thank Ruslan Belikov and Christophe Vérinaud for their insightful discussions, which were highly valuable to this work.

\section*{Disclosures}   

The authors declare no conflicts of interest.

\section*{Data availability}  

No data were generated or analyzed in the presented research.\\

\noindent See Supplement 1 for supporting content.
%%%%%%%%%% If using BibTeX:
\bibliography{biblio}

\begin{thebibliography}{10}
\newcommand{\enquote}[1]{``#1''}

\bibitem{GPI}
B.~Macintosh, J.~R. Graham, P.~Ingraham, \emph{et~al.}, \enquote{First light of the gemini planet imager,} {\protect\JournalTitle{Proceedings of the National Academy of Sciences}} \textbf{111}, 12661–12666 (2014).

\bibitem{sphere}
J.~L. {Beuzit}, A.~{Vigan}, D.~{Mouillet}, \emph{et~al.}, \enquote{{SPHERE: the exoplanet imager for the Very Large Telescope},} {\protect\JournalTitle{Astronomy and Astrophysics}} \textbf{631}, A155 (2019).

\bibitem{faustine_ESO}
F.~Cantalloube, K.~Dohlen, J.~Milli, \emph{et~al.}, \enquote{Peering through sphere images: A glance at contrast limitations,} {\protect\JournalTitle{Published in The Messenger vol. 176}} \textbf{pp. 25-31}, June 2019. (2019).

\bibitem{guyon2010}
O.~Guyon, \enquote{High sensitivity wavefront sensing with a nonlinear curvature wavefront sensor,} {\protect\JournalTitle{Publications of the Astronomical Society of the Pacific}} \textbf{122}, 49--62 (2010).

\bibitem{Chambouleyron_2023}
V.~Chambouleyron, O.~Fauvarque, C.~Plantet, \emph{et~al.}, \enquote{Modeling noise propagation in fourier-filtering wavefront sensing, fundamental limits, and quantitative comparison,} {\protect\JournalTitle{Astronomy and Astrophysics}} \textbf{670}, A153 (2023).

\bibitem{paterson}
C.~Paterson, \enquote{Towards practical wavefront sensing at the fundamental information limit,} {\protect\JournalTitle{Journal of Physics: Conference Series}} \textbf{139}, 012021 (2008).

\bibitem{PhysRevApplied.15.024047}
D.~Bouchet, J.~Dong, D.~Maestre, and T.~Juffmann, \enquote{Fundamental bounds on the precision of classical phase microscopes,} {\protect\JournalTitle{Phys. Rev. Applied}} \textbf{15}, 024047 (2021).

\bibitem{ZeldaMamadou}
M.~N'Diaye, A.~Vigan, K.~Dohlen, \emph{et~al.}, \enquote{{ZELDA, a Zernike wavefront sensor for the fine measurement of quasi-static aberrations in coronagraphic systems: concept studies and results with VLT/SPHERE},} in \emph{{SPIE Astronomical Telescopes + Instrumentation},}  vol. 9909 (Edinburgh, United Kingdom, 2016), p. 99096S.

\bibitem{chambou_Z2WFS}
V.~{Chambouleyron}, O.~{Fauvarque}, J.~F. {Sauvage}, \emph{et~al.}, \enquote{{Variation on a Zernike wavefront sensor theme: Optimal use of photons},} {\protect\JournalTitle{Astronomy and Astrophysics}} \textbf{650}, L8 (2021).

\bibitem{raga}
R.~Ragazzoni, \enquote{Pupil plane wavefront sensing with an oscillating prism,} {\protect\JournalTitle{Journal of Modern Optics}} \textbf{43}, 289--293 (1996).

\bibitem{verinaud_biOedge}
{Vérinaud, C.}, {Héritier, C. T.}, {Kasper, M.}, and {Tallon, M.}, \enquote{The bi–o edge wavefront sensor - how foucault-knife-edge variants can boost extreme adaptive optics,} {\protect\JournalTitle{Astronomy and Astrophysics}} \textbf{682}, A27 (2024).

\bibitem{4QPM}
D.~{Rouan}, P.~{Riaud}, A.~{Boccaletti}, \emph{et~al.}, \enquote{{The Four-Quadrant Phase-Mask Coronagraph. I. Principle},} {\protect\JournalTitle{pasp}} \textbf{112}, 1479--1486 (2000).

\bibitem{Roddier_1997}
F.~Roddier and C.~Roddier, \enquote{Stellar coronograph with phase mask,} {\protect\JournalTitle{Publications of the Astronomical Society of the Pacific}} \textbf{109}, 815 (1997).

\bibitem{fauv_iquad}
{Fauvarque, O.}, {Hutterer, V.}, {Janin-Potiron, P.}, \emph{et~al.}, \enquote{{The ıQuad sensor: a new Fourier-based wave front sensor derived from the 4 quadrants coronagraph},} {\protect\JournalTitle{AO4ELT}}  (2019).

\bibitem{Guyon_2006}
O.~Guyon, E.~A. Pluzhnik, M.~J. Kuchner, \emph{et~al.}, \enquote{Theoretical limits on extrasolar terrestrial planet detection with coronagraphs,} {\protect\JournalTitle{The Astrophysical Journal Supplement Series}} \textbf{167}, 81–99 (2006).

\bibitem{belikov_2021}
R.~{Belikov}, D.~{Sirbu}, J.~B. {Jewell}, \emph{et~al.}, \enquote{{Theoretical performance limits for coronagraphs on obstructed and unobstructed apertures: how much can current designs be improved?}} in \emph{Techniques and Instrumentation for Detection of Exoplanets X,}  vol. 11823 of \emph{Society of Photo-Optical Instrumentation Engineers (SPIE) Conference Series} S.~B. {Shaklan} and G.~J. {Ruane}, eds. (2021), p. 118230W.

\bibitem{vortex_foo}
G.~Foo, D.~M. Palacios, and G.~A. Swartzlander, \enquote{Optical vortex coronagraph,} {\protect\JournalTitle{Opt. Lett.}} \textbf{30}, 3308--3310 (2005).

\bibitem{chambou_SPIE2022}
V.~Chambouleyron, O.~Fauvarque, C.~Plantet, \emph{et~al.}, \enquote{{Optimizing Fourier-filtering WFS to reach sensitivity close to the fundamental limit},} in \emph{Adaptive Optics Systems VIII,}  vol. 12185 L.~Schreiber, D.~Schmidt, and E.~Vernet, eds., International Society for Optics and Photonics (SPIE, 2022), p. 121852T.

\bibitem{landman2022}
R.~Landman, C.~Keller, E.~H. Por, \emph{et~al.}, \enquote{Joint optimization of wavefront sensing and reconstruction with automatic differentiation,}  (2022).

\bibitem{haffert_PIAA_zernike}
S.~Y. Haffert, J.~R. Males, and O.~Guyon, \enquote{Reaching the fundamental sensitivity limit of wavefront sensing on arbitrary apertures with the phase induced amplitude apodized zernike wavefront sensor (piaa-zwfs),}  (2023).

\bibitem{Mawet:09}
D.~Mawet, E.~Serabyn, K.~Liewer, \emph{et~al.}, \enquote{Optical vectorial vortex coronagraphs using liquid crystal polymers: theory, manufacturing and laboratory demonstration,} {\protect\JournalTitle{Opt. Express}} \textbf{17}, 1902--1918 (2009).

\bibitem{Doelman:19}
D.~S. Doelman, F.~F. Auer, M.~J. Escuti, and F.~Snik, \enquote{Simultaneous phase and amplitude aberration sensing with a liquid-crystal vector-zernike phase mask,} {\protect\JournalTitle{Opt. Lett.}} \textbf{44}, 17--20 (2019).

\bibitem{PCS_kasper}
M.~Kasper, N.~Cerpa~Urra, P.~Pathak, \emph{et~al.}, \enquote{Pcs — a roadmap for exoearth imaging with the elt,} {\protect\JournalTitle{Published in The Messenger vol. 182}} \textbf{pp. 38-43}, March 2021. (2021).

\bibitem{salama2024}
M.~Salama, C.~Guthery, V.~Chambouleyron, \emph{et~al.}, \enquote{Keck primary mirror closed-loop segment control using a vector-zernike wavefront sensor,}  (2024).

\bibitem{Milli_2018}
J.~Milli, D.~Mouillet, J.-L. Beuzit, \emph{et~al.}, \enquote{Low wind effect on vlt/sphere: impact, mitigation strategy, and results,} in \emph{Adaptive Optics Systems VI,}  D.~Schmidt, L.~Schreiber, and L.~M. Close, eds. (SPIE, 2018).

\bibitem{bertrou2022}
{Bertrou-Cantou, A.}, {Gendron, E.}, {Rousset, G.}, \emph{et~al.}, \enquote{Confusion in differential piston measurement with the pyramid wavefront sensor,} {\protect\JournalTitle{Astronomy and Astrophysics}} \textbf{658}, A49 (2022).

\bibitem{2024_photonics}
D.~{Sirbu}, R.~{Belikov}, K.~{Fogarty}, \emph{et~al.}, \enquote{{AstroPIC: Architecture options and trades for integrated photonic coronagraphy},} in \emph{American Astronomical Society Meeting Abstracts,}  vol. 243 of \emph{American Astronomical Society Meeting Abstracts} (2024), p. 329.07.

\bibitem{quantum_corono}
N.~Deshler, I.~Ozer, A.~Ashok, and S.~Guha, \enquote{Experimental demonstration of a quantum-optimal coronagraph using spatial mode sorters,}  (2024).

\end{thebibliography}



\section*{Supplementary material}

# Coronagraph-based wavefront sensors for the high Strehl regime: supplemental document

This supplemental document provides the derivation of several important statements used in the main paper titled "Coronagraph-based Wavefront Sensors for the High-Strehl Regime".

## 1. PROJECTION ON THE PISTON MODE

Let's consider the pupil plane EM field : $P(r)e^{i\phi(r)}$. where $P(r)$ is the binary aperture of the telescope. It's projection on the piston mode $s$ (complex-valued scalar) is derived as follow:

$$\rho = < P(r)e^{i\phi(r)} | P(r) > = \frac{1}{\mathcal{N}} \iint_{\mathbb{R}^2} P^2(r) e^{i\phi(r)} dr \tag{S1}$$

where:

$$\mathcal{N} = \iint_{\mathbb{R}^2} P(r) dr \tag{S2}$$

### A. Relation with the Strehl Ratio

The Strehl Ratio (SR) $\mathcal{S}$ can be computed as [1]:

$$\mathcal{S} = \left| \frac{1}{\mathcal{N}} \iint_{\mathbb{R}^2} P(r) e^{i\phi(r)} dr \right|^2 \tag{S3}$$

As $P$ represents the telescope binary aperture we have $P(r) = P(r)^2$, from equations S1 and S3 we obtain the relation: $|\rho| = \sqrt{\mathcal{S}}$.

### B. Value in Maréchal approximation regime

Let's assume the Maréchal approximation: $\sigma_\phi < 1$ radian rms. By taking the Taylor expansion of the phase term, we can rewrite equation S1 and neglects term above $\phi^2$ [1]:

$$\rho = \frac{1}{\mathcal{N}} \iint_{\mathbb{R}^2} P(r) \left[ 1 + i\phi - \frac{\phi^2}{2} + \mathcal{O}(\phi^3) \right] dr \tag{S4}$$

Noting that phase $\phi$ has a zero-mean, we can develop this last equation:

$$\begin{aligned}
\rho &= 1 - \frac{1}{\mathcal{N}} \iint_{\mathbb{R}^2} P(r) \frac{\phi^2}{2} dr \\
&= 1 - \frac{\sigma_\phi^2}{2}
\end{aligned} \tag{S5}$$

Because we assumed $\sigma_\phi < 1$ radian rms, we can infer that $\rho$ is real and positive, thus $\rho = |\rho|$. One can also notice that we retrieve one the well known formula for SR in the case of Maréchal approximation: $\mathcal{S} = \rho^2 = (1 - \frac{\sigma_\phi^2}{2})^2$ [1].

## 2. ENERGY IN REFERENCE WAVE AND SENSITIVITIES

In this appendix section we analyze how the energy carried by the reference wave impact sensitivity to both uniform and photon noise using a simple model of two waves interference. Let's assume that we produce interference between two EM field: reference $\psi_r = \sqrt{I_r}$ with a flat wavefront and a field $\psi_s = \sqrt{I_s} e^{i\phi}$ for which we try to measure the phase $\phi$. The two wave interference formula gives:

$$I(\phi) = I_r + I_s + 2\sqrt{I_r I_s} sin(\phi) \tag{S6}$$

We assume that the energy and the wave $\psi_s$ is fixed and the goal is to analyze how sensitivity evolves when changing energy contained in the reference wave $\psi_r$.

## A. Photon noise

In the case of photon noise, the noise $N$ comes from the signal from the incoherent sum of the two waves:

$$N = \sqrt{I_r + I_s} \tag{S7}$$

and the signal comes from the amplitude of the interferences:

$$S = 2\sqrt{I_r I_s} \tag{S8}$$

Therefore the signal to noise ratio $S/N$ can be written as:

$$S/N = \frac{2\sqrt{I_r I_s}}{\sqrt{I_r + I_s}} \tag{S9}$$

This ratio exhibits two limit cases: when the reference wave carries much less energy than the wave to-be-measured ($I_r \ll I_s$) we have: $S/N \approx 2\sqrt{I_r}$, that is to say that the signal-to-noise ratio is set only by the energy in $I_r$. On the other hand, when energy in the reference wave is much superior we have $S/N \approx 2\sqrt{I_s}$. In that case, the signal-to-noise ratio is set only by the energy in $I_s$. In terms of wavefront sensing described in this paper, the latter case correspond to the one for small phase regime. To analyze intermediate cases, we define the energy in the reference wave as a ratio of the energy in the wave to-be-measured as $I_r = \eta I_s$ (where $\eta \in \mathbb{R}^+$). In that case we can rewrite equation S9:

$$S/N = 2\sqrt{I_s}\frac{\sqrt{\eta}}{\sqrt{1+\eta}} \tag{S10}$$

As we can see on figure S1, once $\eta = 4$ ($I_r = 4I_s$) the $S/N$ is already reaching 90% of its maximum value. Passed this value, most of the energy added doesn't bring significant increase in the $S/N$ value.

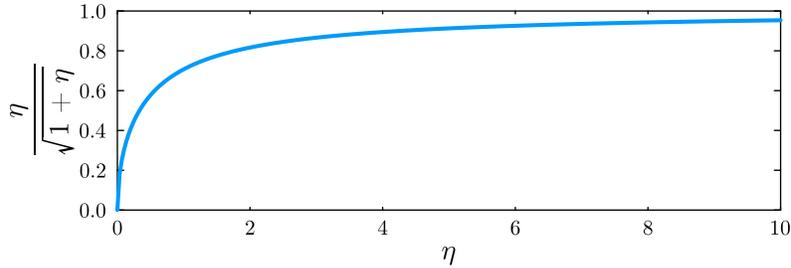

**Fig. S1.** Graphical representation of the function $f(\eta) = \frac{\sqrt{\eta}}{\sqrt{1+\eta}}$

## B. Detector noise

In the case of a detector noise defined by its standard deviation $\sigma$, the $S/N$ is written:

$$S/N = \frac{2\sqrt{I_r I_s}}{\sigma} \tag{S11}$$

In that case, for a fixed $I_s$, increasing $I_r$ will simply always increases signal-to-noise ratio.

## 3. IMPACT OF PISTON LEAKAGE ON THE IDEAL WFS SENSITIVITIES

This section aims at quantifying the impact of piston leakage on the iWFS sensitivities. By piston leakage, we refer to a portion of the energy carried by the EM piston mode not being phase-shifted by the iWFS. Let's assume the small phase regime. In that case, we recall the iWFS operation on the EM field:

$$\mathcal{W}[\psi] = \mathcal{W}[P + iP\phi] = iP + iP\phi \tag{S12}$$

Now, let's assume a piston leakage described by a real scalar $\nu \in [0, 1]$ which represents the portion of the energy of the piston mode being effectively phase-shifted by $\pi/2$. It is then posssible to write:

$$\mathcal{W}[\psi] = \mathcal{W}[P + iP\phi] = i\sqrt{\nu}P + \sqrt{1-\nu}P + iP\phi \tag{S13}$$

where $\sqrt{\nu}P$ the part of the piston mode amplitude being phase-shifted while $\sqrt{1-\nu}P$ represents the part which is not phase-shifted. By developing the associated intensities and neglecting the terms in $\phi^2$:

$$I(\phi) = |\mathcal{W}[\psi]|^2 = \nu P + (1-\nu)P + 2\nu P\phi = P + 2\sqrt{\nu}P\phi \tag{S14}$$

Using equations (3) and (5) in the main paper, one can compute the associated sensitivities which simply take the form $s = s_\gamma = 2\sqrt{\nu}$. From this analysis, it is possible to conclude that a small piston leakage doesn't impact drastically the iWFS sensitivities, as a few percent piston leakage will simply reduce the sensitivities by also a few percent.

**REFERENCES**

1. V. N. Mahajan, "Strehl ratio for primary aberrations: some analytical results for circular and annular pupils," J. Opt. Soc. Am. **72**, 1258–1266 (1982).



\end{document}